\documentclass[12pt,preprint]{aastex}
\slugcomment{Not to appear in Nonlearned J., 45.}

\shorttitle{Asymmetric Damped Ly$\alpha$ and Ly$\beta$ Profiles}

\shortauthors{Lee}

\begin{document}
\title{Asymmetric Absorption Profiles of Ly$\alpha$ and Ly$\beta$ in Damped Lyman Alpha Systems
}

\author{Hee-Won Lee} 
\affil{Department of Astronomy and Space Science, Sejong University,
Seoul, 143-747, Korea \\
{\rm hwlee@sejong.ac.kr}}

\begin{abstract}
Damped Ly$\alpha$ systems (DLAs) observed in the quasar spectra are 
characterized by a high neutral hydrogen column density
$N_{HI}>2\times 10^{20}{\rm\ cm^{-2}}$. The absorption wing profiles are often fitted 
using the Voigt function due to the fact that
the scattering cross section near resonant line center is approximately described 
by the Lorentzian function. Since a hydrogen atom has infinitely many
$p$ states that participate in the electric dipole interaction, the cross section starts to deviate from the Lorentzian in an
asymmetric way in the line wing regions.
We investigate this asymmetry in the absorption line profiles around Ly$\alpha$ 
and Ly$\beta$ as a function 
of the neutral hydrogen column density $N_{HI}$. In terms of $\Delta\lambda\equiv\lambda-\lambda_\alpha$ 
we expand the Kramers-Heisenberg 
formula around Ly$\alpha$ to find $\sigma(\lambda)\simeq (0.5 f_{12})^2\sigma_{T}
(\Delta\lambda/\lambda_\alpha)^{-2}[1+3.792 (\Delta\lambda/\lambda_\alpha)]$, where $f_{12}$ and $\sigma_T$
are the oscillator strength of Ly$\alpha$ and the Thomson scattering cross section, respectively. 
In terms of $\Delta\lambda_2\equiv\lambda-\lambda_\beta$ in the vicinity of Ly$\beta$,
 the total scattering cross section, given as the sum of cross sections 
for Rayleigh and Raman scattering, is
shown to be $\sigma(\lambda)\simeq \sigma_T(0.5f_{13})^2(1+R_0)
(\Delta\lambda_2/\lambda_\beta)^{-2}[1-24.68(\Delta\lambda_2/\lambda_\beta)]$ with $f_{13}$ and the factor $R_0=0.1342$ 
being the oscillator strength for
Ly$\beta$ and the ratio of Raman cross section to Rayleigh cross section, respectively.
A redward asymmetry develops around Ly$\alpha$ whereas 
a blue asymmetry is obtained for Ly$\beta$. 
The absorption center shifts are found to be almost proportional 
to the neutral hydrogen column density.  

\end{abstract}

\section{Introduction}

Hydrogen is the most abundant element in the universe and therefore 
is an important probe for the physical conditions of
intergalactic medium throughout the observable universe. High resolution spectroscopy of quasars 
shows a large number of absorption lines blueward of Ly$\alpha$ mostly due to neutral hydrogen 
components in intergalactic medium intervening along the line of sight. 
These quasar absorption systems due to hydrogen are classified according to the
\ion{H}{1} column densities. 
Quasar absorption systems with a neutral column density not exceeding $N_{HI}\simeq10^{17}{\rm\ cm^{-2}}$ 
constitute the Ly$\alpha$ 
forest which is attributed to the residual hydrogen atoms contained in intergalactic filamentary structures 
that are highly ionized  (e.g. Rauch 1998, Meiksin 2009, Kim et al. 2011). 
The absorption systems associated with a neutral hydrogen
column density in excess of $N_{HI}=10^{20.3}{\rm\ cm^{-2}}$ are called 
the damped Lyman alpha systems (DLAs), which are distinguished from other quasar absorption
systems in that they are dominantly neutral (e.g. Wolfe et al. 1986, 2005).

A catalogue of 322 DLAs was provided by Curran et al. (2002), and recently 
721 DLAs 
are listed from the Sloan Digital Sky Survey Data Release 7 (Khare et al. 2012). 
The exact nature of the DLAs is still controversial,
but they are believed to dominate the neutral gas content in the universe, providing raw material for star formation during
most of the time from the reionization era to the present (e.g.  Wolfe et al. 2005, Prochaska, Herbert-Fort \& Wolfe 2005). 
DLAs are also important to trace
the chemical evolution of the universe by carefully measuring the metal abundance as a function 
of redshift (e.g. Calura, Matteucci \& Vladilo 2003, 
Rafelski et al. 2012). Accurate atomic physics for Ly$\alpha$ is essential
to obtain a reliable estimate of metallicity of a DLA.

Recently, Kulkarni et al. (2012) reported a discovery of ''super-damped'' Lyman-alpha absorber at
$z_{abs}=2.2068$ toward the QSO Q1135-0010. Their profile fit to the DLA showed a high neutral hydrogen column density
of $N_{HI}=10^{22.05}{\rm\ cm^{-2}}$. A little higher column density of $N_{HI}=10^{22.10}{\rm\ cm^{-2}}$
was reported for the same object by Noterdaeme et al. (2012), who also suggested a significant star formation rate of 
$25{\rm\ M_\odot\ yr^{-1}}$ based on the strength of H$\alpha$.

Damped Ly$\alpha$ absorption profiles are analyzed
using the Voigt function which is defined as a convolution of a Lorentzian function with a Gaussian function 
(e.g. Rybicki \& Lightman 1979). When $N_{HI}$ is very large, the absorption at the core part is almost complete. Contributing
only to the core part of the Voigt function, the Gaussian function does not play an important role in the analysis 
of high column density systems. 
In the wing part the Voigt function coincides with the Lorentzian, and in this case the damping term in the denominator
of the Lorentzian is quite negligible. Therefore, the wing profile is essentially proportional to $\Delta\lambda^{-2}$, 
where $\Delta\lambda=\lambda-\lambda_\alpha$ is the difference in wavelength  
from the Ly$\alpha$ line center wavelength $\lambda_\alpha=1215.671{\rm\ \AA}$.

The Lorentzian function, invoked to describe the wing parts
of the absorption profile, is obtained  when the scattering atom is regarded 
as a two-level atom.  
Instead of being a two-level system, the hydrogen atom has infinitely many 
energy levels, and therefore the scattering cross section is expected to deviate from the Lorentzian 
function that is symmetric with respect to the line center
wavelength. 
Peebles (1993) discussed the resonance line shape that deviates from the simple Lorentzian and mentioned the marginal possibility 
of detecting this deviation in the absorption line profiles of DLA systems (e.g. Peebles 1993, p. 573). The formula for the scattering
cross section he introduced is derived in a heuristic way to illustrate the behavior of the Lorentzian profile near line resonance and 
the classical $\omega^4$-dependence in the low energy regime. However, this formula is inaccurate because it fails to
include the contributions from the infinitely many $p$ states
that participate in the electric dipole interaction.

 The exact scattering cross section is computed by summing all the probability
amplitudes contributed from the infinitely many bound and free $p$ states.  The result is summarized 
in the Kramers-Heisenberg formula (e.g. Bethe \& Salpeter 1967, Sakurai 1967).
An expansion of the Kramers-Heisenberg formula around Ly$\alpha$ in frequency space was given by Lee (2003), 
in which the redward asymmetry of the scattering cross section around Ly$\alpha$ was briefly illustrated. 
However, spectroscopy is often presented in wavelength space rather than in frequency
space so that it will also be useful to express the Kramers-Heisenberg formula in wavelength space.
In this paper, we present the same expansion in wavelength space and quantify the redward shift of the center
wavelength as a function of $N_{HI}$. In addition,
we also expand the Kramers-Heisenberg formula in the vicinity of Ly$\beta$ in order to investigate the asymmetry
around Ly$\beta$.

In the case of Ly$\beta$, there is an additional scattering channel, which results from  radiative de-excitation
into the $2s$ state re-emitting an H$\alpha$ line photon. This inelastic or Raman scattering branch
is proposed to be important in the formation of broad H$\alpha$ wings observed in young planetary nebulae
and symbiotic stars (e.g. Isliker, Nussbaumer \& Vogel 1989, Lee 2000, Schmid 1989). In this paper, we show that the cross 
section around Ly$\beta$ is asymmetric
blueward, which is in high contrast with the behavior around Ly$\alpha$. 
A brief discussion on the observational consequences is presented.

\section{Calculation}

\subsection{Scattering cross section around Ly$\alpha$ in wavelength space}

The scattering cross section is given by the Kramers-Heisenberg formula that 
is obtained from a second-order time dependent perturbation theory (e.g.
Sakurai 1967, Merzbacher 1970).
In terms of the matrix elements of the dipole operator, 
the Kramers-Heisenberg formula can be written as
\begin{eqnarray}
{d\sigma\over d\Omega}&=& {r_0^2 m_e^2 \over \hbar^2} \left|
\sum_I \omega\omega_{I1}\left({{({\bf r}\cdot {\bf e}^{(\alpha')})_{1I}
({\bf r}\cdot {\bf e}^{(\alpha)})_{I1}}\over{\omega_{I1}-\omega}} \right.\right.
\nonumber \\ 
&-& \left.\left.  {{({\bf r}\cdot {\bf e}^{(\alpha)})_{1I}
({\bf r}\cdot {\bf e}^{(\alpha')})_{I1}}\over{\omega_{I1}+\omega}}
\right)
\right|^2,
\label{kr_hei} 
\end{eqnarray}
where $m_e$ is the electron mass and $r_0=e^2/m_e c^2$ is the classical electron 
radius. The polarization vectors of incident and scattered radiation are denoted by ${\bf e}^{(\alpha)}$ and ${\bf e}^{(\alpha')}$, respectively.
Here, $\omega_{I1}$ is the angular frequency between the intermediate state $I$ 
and the ground $1s$ state. The intermediate state $I$ that participates 
in the electric dipole interaction of Lyman photons consists of bound
$np$ states and free $n'p$ states.
In the atomic units adopted in this work, the bound $np$ state has the energy
eigenvalue $E_n = -1/(2 n^2)$ and correspondingly $\omega_{I1}=\omega_{n1}=(1-n^{-2})/2$.
Similarly, for the $n'p$ state, $E_{n'}=1/(2n'^2)$ and $\omega_{I1}=\omega_{n'1}=(1+n'^{-2})/2$.
The term denoted by $({\bf r}\cdot {\bf e}^{(\alpha)})_{I1}$ represents
the matrix element of the position
operator between the intermediate state $I$ and the ground $1s$ state.

Being a two-body system, the hydrogen atom admits analytically closed 
expressions of the wavefunctions, which enables one to compute explicitly 
the matrix elements of the dipole operator in terms of the confluent
hypergeometric function.  A typical matrix element 
$({\bf r}\cdot \epsilon^{(\alpha)})_{I1}$ for an intermediate state with $I=|np,m>$
is explicitly written as
\begin{equation}
({\bf r}\cdot {\bf e}^{(\alpha)})_{I1}=
\int_0^{2\pi}\int_0^\pi\int_0^\infty [R_{nl}(r)Y_l^m(\theta,\phi)]^* ({\bf r}\cdot 
{\bf e}^{(\alpha)})
R_{10}(r)Y_0^0(\theta,\phi) r^2 dr \sin\theta d\theta d\phi,
\end{equation}
where $Y_l^m(\theta,\phi)$ is a spherical harmonic function and $R_{nl}(r)$ is
the radial wavefunction given by an associated Laguerre function. 
The angular integration followed by summing over magnetic substates
$m=\pm1, 0$ of each $np$ state and averaging over polarization states of incident
and outgoing radiation results in a numerical factor of ${8\pi\over 3}$, which is discussed
in more detail in Appendix A.

The radial matrix elements of the dipole operators 
$<r>_{n1}$ and $<r>_{n',1}$ are readily found in many textbooks 
on quantum mechanics (e.g. Berestetski, Lifshitz \& Pitaevskii 1971, 
Bethe \& Salpeter 1967). The radial matrix elements for the bound $np$ states 
are given by
\begin{equation}
<r>_{1n}=<1s| r | np> =\left[ {2^{8}n^7 (n-1)^{2n-5}}\over
{(n+1)^{2n+5}} \right]^{1\over2} a_B,
\end{equation}
where $a_B=\hbar^2/ me^2=0.5292{\rm\ \AA}$ is the Bohr radius.
For the continuum $n'p$ states, the corresponding values are given by
\begin{equation}
<r>_{1n'} = <1s| r| n'p>
=\left[ {2^{8}(n')^7 \exp[-4n'\tan^{-1}(1/n')]}\over
{[(n')^2+1]^{5}[1-\exp(-2\pi n')]} \right]^{1\over2}a_B,
\end{equation}
which is obtained through analytic continuation into the complex
plane (Bethe \& Salpeter 1967, Saslow \& Mills 1969).

Lee (2003) provided the expansion of the Kramers-Heisenberg formula for Rayleigh scattering 
in the vicinity of Ly$\alpha$ in terms of $\Delta\omega=\omega-\omega_{21}$. 
The result is summarized as
\begin{eqnarray}
\sigma(\omega) &=& {\sigma_T}
\left({f_{12}\over 2}\right)^2\left( {\omega_{21}\over\Delta\omega} \right)^2
\Bigg|1+a_1\left({\Delta\omega\over\omega_{21}}\right)   \nonumber \\
&&+a_2\left({\Delta\omega\over\omega_{21}}\right)^2 
  +\cdots \Bigg|^2, 
\end{eqnarray}
where $\sigma_T = 8\pi r_0^2/3=0.6652\times 10^{-24}{\rm\ cm^2}$ is the Thomson scattering cross section and  
$f_{12}=0.4162$ is the oscillator strength for the Ly$\alpha$ transition. 
The coefficients $a_k$  were numerically computed by Lee (2003), who gave
\begin{equation}
a_1 =  -8.961\times 10^{-1},\ a_2 =-1.222\times 10^1 .
\end{equation}
In Table~1, we show these coefficients up to $a_5$.

 Up to the first order of $\Delta\omega/\omega_{12}$ 
the scattering cross section can be expressed as
\begin{eqnarray}
\sigma(\omega)  &\simeq& 
\sigma_T \left({f_{12}\over2}\right)^2 
\left({\omega_{12}\over\Delta\omega}\right)^2
\left(1+2a_1{\Delta\omega\over\omega_{21}}\right)
\nonumber \\
&=&
\sigma_\alpha \left({\omega_{12}\over\Delta\omega}\right)^2
\left(1-1.792{\Delta\omega\over\omega_{21}}\right).
\label{lyaeq1}
\end{eqnarray}
Here, we introduce the characteristic cross section $\sigma_\alpha$ 
for Ly$\alpha$ defined as $\sigma_\alpha\equiv \sigma_T(f_{12}/2)^2=2.880\times
10^{-26}{\rm\ cm^2}$.
The coefficient $a_1$, being less than zero, is responsible 
for the redward asymmetric 
deviation of the scattering cross section in frequency space.

Astronomical spectroscopy is often presented in wavelength space, which makes 
it necessary to express the scattering cross section in terms of $\Delta\lambda=\lambda-\lambda_\alpha$,
the difference in wavelength from the Ly$\alpha$ line center.
From the following relation
\begin{equation}
{\Delta\omega\over\omega_\alpha} =-{\Delta\lambda\over\lambda}
=\sum_{n=1}^\infty \left(-{\Delta\lambda\over\lambda_\alpha}\right)^n,
\label{convert}
\end{equation}
we may notice that 
expansion in wavelength space will yield different coefficients 
from those obtained in expansion in frequency space.
Substituting Eq.~(\ref{convert})  into Eq.~(\ref{lyaeq1}),  we obtain 
\begin{eqnarray}
\sigma(\lambda)&\simeq&\sigma_T \left({f_{12}\over2}\right)^2 \left({\lambda_{\alpha}\over\Delta\lambda}\right)^2
\Bigg[1+2(1-a_1)\left({\Delta\lambda\over\lambda_{\alpha}}\right)
\nonumber \\
&+&(1-2a_1+2a_2-a_1^2)
\left({\Delta\lambda\over\lambda_{\alpha}}\right)^2
\nonumber\\
&-&2(a_3+a_1a_2)
\left({\Delta\lambda\over\lambda_{\alpha}}\right)^3+\cdots\Bigg]
\nonumber \\
&=&\sigma_\alpha \left({\lambda_{\alpha}\over\Delta\lambda}\right)^2
\Bigg[1+3.792\left({\Delta\lambda\over\lambda_{\alpha}}\right)
\nonumber\\
&-&20.84 \left({\Delta\lambda\over\lambda_{\alpha}}\right)^2
+83.14\left({\Delta\lambda\over\lambda_{\alpha}}\right)^3+\cdots \Bigg].
\label{lyaeq2}
\end{eqnarray}
Thus, up to first order, we have 
\begin{equation}
\sigma(\lambda)\simeq \sigma_\alpha  \left({\lambda_{\alpha}\over\Delta\lambda}\right)^2
\Bigg[1+3.792\left({\Delta\lambda\over\lambda_{\alpha}}\right)\Bigg],
\label{lya_wv}
\end{equation}
from which it is found that the coefficient 3.792 is significantly 
different from the coefficient $-1.792$ in frequency space.

The red asymmetry in the Ly$\alpha$ scattering cross section can be explained by noting the
denominator $\omega_{n1}-\omega$ in Eq.~(1), where the dominant contribution comes from $n=2$. 
The contributions from all the excited states
with $n>2$ interfere positively with the dominant contribution from $n=2$ for photons
on the red side whereas the interference is negative for photons on the blue side.
Therefore the red wing is strengthened relative to the pure Lorentzian profile
by the positive interference of scattering from all other levels.

\subsection{Rayleigh scattering cross section around Ly$\beta$}

In a similar way, we define
the difference in frequency from Ly$\beta$ by
\begin{equation}
\Delta\omega_2 \equiv \omega-\omega_{31},
\end{equation}
and expand the Kramers-Heisenberg formula in terms of $\Delta\omega_2/\omega_{31}$.
We note that  the terms appearing in the Kramers-Heisenberg formula include
\begin{equation}
{\omega\omega_{31} \over \omega_{31}-\omega}=-\omega_{31}
\left(1+{\omega_{31}\over \Delta\omega_2} \right),
\end{equation}
and 
\begin{equation}
{\omega \over \omega_{n1}-\omega} =
{\omega_{31}\over\omega_{n1}-\omega_{31}}
\left[ 1 +{\omega_{n1}\over \omega_{31}}
\sum_{k=1}^\infty \left( {\Delta\omega_2\over \omega_{n1}-\omega_{31}} \right)^k \right].
\end{equation}
For $n\ge 2$, the second term in the summation in Eq.~(\ref{kr_hei}) can be written as
\begin{equation}
{\omega \over \omega_{n1}+\omega} = 
{\omega_{31}\over \omega_{n1}+\omega_{31}}
\left[ 1 -{\omega_{n1}\over \omega_{31}}
\sum_{k=1}^\infty \left( {-\Delta\omega_2\over\omega_{n1}+\omega_{31}} \right)^k
\right].
\end{equation}
Similar expressions for the continuum states are obtained 
in a straightforward manner.

We expand the Kramers-Heisenberg formula in frequency space for Rayleigh scattering near Ly$\beta$, which is written as
\begin{eqnarray}
\sigma^{Ray}(\omega) &=& {\sigma_T\over 9} \left({\omega_{31}\over\Delta\omega_2}\right)^2
\left| \omega_{31}<r>_{13}^2 +\omega_{31}<r>_{13}^2 
\left({\Delta\omega_2\over \omega_{31}}\right) \right. \nonumber \\
&-& \sum_{n\neq3}
{\Delta\omega_2\omega_{n1}\over\omega_{n1}-\omega_{31}}
\left[ 1 +{\omega_{n1}\over \omega_{31}}
\sum_{k=1}^\infty \left( {\Delta\omega_2\over \omega_{n1}-\omega_{31}} \right)^k \right]
<r>_{n1}^2 
\nonumber \\
&+& \sum_{n\neq 3}
{\Delta\omega_2\omega_{n1}\over \omega_{n1}+\omega_{31}}
\left[ 1 -{\omega_{n1}\over \omega_{31}}
\sum_{k=1}^\infty \left( {-\Delta\omega_2\over\omega_{n1}+\omega_{31}} \right)^k
\right] <r>_{n1}^2  \nonumber \\
&-& \int_0^\infty dn'
{\Delta\omega_2\omega_{n'1}\over\omega_{n'1}-\omega_{31}}
\left[ 1 +{\omega_{n'1}\over \omega_{31}}
\sum_{k=1}^\infty \left( {\Delta\omega_2\over \omega_{n'1}-\omega_{31}} \right)^k \right]
<r>_{n'1}^2 
\nonumber \\
&+& \left. \int_0^\infty dn'
{\Delta\omega_2\omega_{n'1}\over \omega_{n'1}+\omega_{31}}
\left[ 1 -{\omega_{n'1}\over \omega_{31}}
\sum_{k=1}^\infty \left( {-\Delta\omega_2\over\omega_{n'1}+\omega_{31}} \right)^k
\right] <r>_{n'1}^2 \right|^2.
\label{eq_lyb}
\end{eqnarray}
Here,  angular integration has been performed and the atomic unit system 
is adopted, in which the Bohr radius $a_B=1$. 

An algebraic rearrangement of Eq.~(\ref{eq_lyb}) can be made to express $\sigma^{Ray}(\omega)$ as 
\begin{eqnarray}
\sigma^{Ray}(\omega) &=& \sigma_T\left( {\omega_{31}\over\Delta\omega_2} \right)^2
\Bigg|B_0+B_1\left({\Delta\omega_2\over\omega_{31}}\right)  
\nonumber \\
&& +B_2\left({\Delta\omega_2\over\omega_{31}}\right)^2 
  +\cdots \Bigg|^2.
\end{eqnarray}
The coefficients $B_0$ and $B_1$ are determined through following relations;
\begin{eqnarray}
B_0 &=& {\omega_{31}\over 3}<r>_{13}^2 = f_{13}/2 =0.03955\nonumber \\
B_1 &=& {1\over2}\omega_{31}<r>_{13}^2 
-{2\over3}\sum_{n\neq3}{\omega_{31}^2\omega_{n1}
\over\omega_{n1}^2-\omega_{31}^2}<r>_{1n}^2   
\nonumber \\
&-& {2\over3} \int_0^\infty dn' {\omega_{31}^2\omega_{n'1}
\over\omega_{n'1}^2-\omega_{31}^2}<r>_{n'1}^2 =0.6414,
\end{eqnarray}
where $f_{13}=0.07910$ is the oscillator strength for the Ly$\beta$ transition.
The coefficients $B_k$ for $k\ge 2$ are given as
\begin{eqnarray}
B_k &=&\left({-1\over 2}\right)^k {\omega_{31}\over3}<r>_{13}^2
\nonumber \\
&-&{1\over3}\sum_{n\neq3}^\infty {\omega_{n1}^2\over\omega_{31}}
\left[
\left({\omega_{31}\over\omega_{n1}-\omega_{31}}\right)^k
-\left({-\omega_{31}\over\omega_{n1}+\omega_{31}}\right)^k
\right]<r>_{n1}^2 \nonumber \\
&-&{1\over3}\int_{0}^\infty dn' {\omega_{n'1}^2\over\omega_{31}}
\left[
\left({\omega_{31}\over\omega_{n'1}-\omega_{31}}\right)^k
-\left({-\omega_{31}\over\omega_{n'1}+\omega_{31}}\right)^k
\right]<r>_{n'1}^2. 
\end{eqnarray}

The numerical values of the coefficients
 up to $k=5$ are computed as follows;
\begin{eqnarray}
b_1=B_1/B_0 & = & 1.621\times 10^1  \nonumber \\
b_2=B_2/B_0 & = & -4.299\times 10^2   \nonumber \\
b_3=B_3/B_0 & = & -2.176\times 10^3   \nonumber \\
b_4=B_4/B_0 & = & -6.005\times 10^4   \nonumber \\
b_5=B_5/B_0 & = & -8.414\times 10^5   .
\end{eqnarray}
In particular, the coefficient $b_1$ is positive due to the predominant contribution 
from the $2p$ state.

Therefore, up to the first order approximation in $\Delta\omega_2/\omega_{31}$, 
the Rayleigh scattering cross
section around Ly$\beta$ is given by 
\begin{eqnarray}
\sigma^{Ray}(\omega) &\simeq& \sigma_T\left({f_{13}\over 2}\right)^2 
\left({\omega_{13}\over\Delta\omega_2}\right)^2 \left(1+2b_1{\Delta\omega_2\over\omega_{31}}\right)
\nonumber \\
&\simeq& \sigma_T \left({f_{13}\over2}\right)^2 \left({\omega_{13}\over\Delta\omega_2}\right)^2
\left(1+32.42{\Delta\omega_2\over\omega_{31}}\right).
\label{lybeq1}
\end{eqnarray}

In wavelength space, the Rayleigh scattering cross section in the vicinity of Ly$\beta$ with the line center $\lambda_\beta=1025.722 {\rm\ \AA}$
can be expanded as
\begin{eqnarray}
\sigma^{Ray}(\lambda)  & \simeq& \sigma_T \left({f_{13}\over2}\right)^2 \left({\lambda_{\beta}\over\Delta\lambda_2}\right)^2
\Bigg[1+2(1-b_1)\left({\Delta\lambda_2\over\lambda_{\beta}}\right)
\nonumber \\
&+&(1-2b_1+2b_2-b_1^2)
\left({\Delta\lambda_2\over\lambda_{\beta}}\right)^2
\nonumber\\
&-&2(b_3+b_1b_2)
\left({\Delta\lambda_2\over\lambda_{\beta}}\right)^3+\cdots\Bigg]
\nonumber \\
&=&\sigma_T\left({f_{13}\over2}\right)^2 \left({\lambda_{\beta}\over\Delta\lambda_2}\right)^2
\Bigg[1-31.64\left({\Delta\lambda_2\over\lambda_{\beta}}\right)
\nonumber\\
&-&5.970\times10^2 \left({\Delta\lambda_2\over\lambda_{\beta}}\right)^2
+1.792\times10^4\left({\Delta\lambda_2\over\lambda_{\beta}}\right)^3\cdots \Bigg],
\label{lybeq2}
\end{eqnarray}
where $\Delta\lambda_2\equiv\lambda-\lambda_\beta$ is the wavelength deviation from the Ly$\beta$ center. 

The negative value of the coefficient $2(1-b_1)$
implies that the Rayleigh scattering cross section near Ly$\beta$ 
is asymmetric to the blue of Ly$\beta$, which is
in high contrast with the behavior around Ly$\alpha$.
In the case of Ly$\beta$, the main contributor to the Kramers-Heisenberg formula
is the $3p$ state and the residual contribution comes from the $2p$ state and
all the $p$ states lying higher than the $3p$ state. 
The contribution to $\sigma(\lambda)$ of a given $np$ or $n'p$ state   
is measured roughly by the oscillator strength inversely weighted 
by the energy difference from Ly$\beta$. Therefore, the contribution of the $2p$
state is more important than that from all the $p$ states lying higher than the
$3p$ state. Hence, in the case of Ly$\beta$, the $n=2$ contribution has lower
frequency and the interference with the principal $n=3$ scattering contribution
is negative on the red side and positive on the blue side, which explains the
blue asymmetric scattering cross section.

\subsection{Raman scattering cross section around Ly$\beta$}

Interaction with a hydrogen atom of electromagnetic radiation around 
Ly$\beta$ has another channel, which is 
inelastic or Raman scattering.
The scattering hydrogen atom de-excites into the $2s$ state re-emitting 
an H$\alpha$ photon into another line of sight,
which provides an important contribution to the absorption profile around Ly$\beta$. 
 The astrophysical
importance of Raman scattering can be appreciated in the emission features at
6830 \AA\ and 7088 \AA\ that appear in the spectra of about a half of symbiotic stars.
These are formed through Raman conversion of the resonance doublet  
\ion{O}{6}$\lambda\lambda$ 1032, 1038 (Schmid 1989,
Nussbaumer, Schmid \& Vogel 1989). Another example of Raman scattering by
atomic hydrogen is provided by far UV \ion{He}{2} emission lines
in symbiotic stars and young planetary nebulae (e.g. Birriel 2004, Lee et al. 2006, Lee 2012).
It has also been proposed that broad 
H$\alpha$ wings often found
in planetary nebulae and symbiotic stars are formed through Raman scattering 
of far UV continuum around Ly$\beta$ 
(e.g. Lee 2000, Arrieta \& Torres-Peimbert 2003). 

As is illustrated in Sakurai (1967), the term corresponding to the 'seagull graph' is 
absent in the Kramers-Heisenberg formula for the case of Raman scattering.
This difference allows an alternate expression of the Kramers-Heisenberg formula 
given in terms of the matrix elements of the momentum operator 
(see also Saslow \& Mills 1969, Lee \& Lee 1997). 
In a manner analogous to what is illustrated in Appendix A, taking angular integrations,
summing over magentic substates and averaging over polarizations of incident and 
outgoing radiation, we arrive at an
explicit expression of the Raman cross section given by
\begin{eqnarray}
\sigma^{Ram}(\omega) &=&{\sigma_T\over9}\left({\omega'\over\omega}\right)
\Bigg|\sum_{n=3}^\infty <p>_{n1}<p>_{n2}
\left({1\over \omega_{n1}-\omega}+{1\over \omega_{n1}+\omega'} \right)
\nonumber \\
&+&\int_0^\infty dn' <p>_{n'1}<p>_{n'2}
\left({1\over \omega_{n'1}-\omega}+{1\over \omega_{n'1}+\omega'} \right)
\Bigg|^2.
\label{raman_kh}
\end{eqnarray}
Here $\omega'=\omega-\omega_{21}$ is the angular frequency of the Raman scattered radiation. 
The matrix element $<p>_{n1}$ associated with the momentum operator between the $np$ 
and the $1s$ states is given by 
\begin{equation}
<p>_{n1}=\int_0^\infty R_{n1}(r)
\left[{d\over d r}R_{10}(r)\right]
 r^2 dr=
\left[{ 2^6 n^3 (n-1)^{2n-3}\over (n+1)^{2n+3}}
\right]^{1/2}.
\end{equation}
Here, an atomic unit system is adopted and the reality of the radial wavefunctions is
noted.

The  matrix element $<p>_{n2}$ corresponds to the transition 
between the $np$ and $2s$ states, which is explicitly given by
\begin{equation}
<p>_{n2}=\left[{ 2^{11} n^3 (n^2-1)(n-2)^{2n-4}\over (n+2)^{2n+4}}
\right]^{1/2}.
\end{equation}
The contribution from the continuum $n'p$ states is obtained by considering the matrix elements of the momentum operator given by
\begin{eqnarray}
<p>_{n'1} &=&\left[{ 2^6 n'^3 e^{-4n'\tan^{-1}{1\over n'}}\over (n'^2+1)^3(1-e^{-2\pi n'})}
\right]^{1/2}
\nonumber \\
<p>_{n'2} &=&\left[{ 2^{11} n'^3(n'^2+1) e^{-4n'\tan^{-1}{2\over n'}}\over (n'^2+4)^4(1-e^{-2\pi n'})}
\right]^{1/2}.
\end{eqnarray}
Due to the vanishing matrix element $<p>_{n2}$ for $n=2$,  
the sum in Eq.~(\ref{raman_kh}) begins from $n=3$ for the bound 
$np$ states, which implies that the $2p$ state does not contribute to the cross
section for Raman scattering around Ly$\beta$. This is 
decisively important to the behavior of the cross section, as we
discuss later in more detail.

The terms involving angular frequencies can be rearranged using the following relation
\begin{eqnarray}
{\omega'\over\omega} &=& {\omega_{31}-\omega_{21}+\Delta\omega \over
\omega_{31}+\Delta\omega}
\nonumber \\
&=&{\omega_{32}\over \omega_{31}}-
{\omega_{21}\over\omega_{31}} \sum_{k=1}^\infty \left({-\Delta\omega\over \omega_{31}}\right)^k,
\end{eqnarray}
where $\omega_{32}=\omega_{31}-\omega_{21}$ is the angular frequency for
H$\alpha$.
Use is also made of the following relations
\begin{eqnarray}
{1\over \omega_{n1}-\omega} &=&{1\over\omega_{31}}
\sum_{k=0}^\infty \left( {\Delta\omega\over \omega_{31}} \right)^k
\left({\omega_{31}\over \omega_{n1}-\omega_{31}}\right)^{k+1} 
\quad {\rm for}\quad {n\ge4}
\nonumber \\
{1\over \omega_{n1}+\omega'}&=&-{1\over\omega_{31}}
\sum_{k=0}^\infty \left( {\Delta\omega\over \omega_{31}} \right)^k
\left({-\omega_{31}\over \omega_{n1}+\omega_{31}-\omega_{21}}\right)^{k+1}.
\end{eqnarray}

The Kramers-Heisenberg formula for Raman scattering near Ly$\beta$ can be 
expanded in frequency space as follows
\begin{equation}
\sigma^{Ram}(\omega)=\sigma_T\left({\omega_{31}\over\Delta\omega}\right)^2
\left({\omega'\over\omega}\right)
\Bigg| C_0+C_1(\Delta\omega/\omega_{31})+C_2(\Delta\omega/\omega_{31})^2+\cdots\Bigg|^2,
\end{equation}
where the coefficients $C_k$
 are given by
\begin{eqnarray}
C_{0} &=&-{<p>_{32}<p>_{31}\over 3\omega_{31}}
=-{1\over2}
\left({\omega_{32}\over\omega_{31}}\right)^{1/2}(f_{13}f_{2s,3p})^{1/2}
\nonumber\\
&=&-3^4\cdot 2^{1/2}\cdot 5^{-5}
=-0.03666
\nonumber \\
C_1&=&{<p>_{32}<p>_{31}\over 3(2\omega_{31}-\omega_{21})}+{1\over3}\sum_{n\ge 4}
<p>_{n2}<p>_{n1}\left( {1\over \omega_{n1}-\omega_{31}}+
{1\over \omega_{n1}+\omega_{31}-\omega_{21}} \right)
\nonumber\\
&+&{1\over3}\int_0^\infty dn' <p>_{n'2}<p>_{n'1}\left( {1\over \omega_{n'1}-\omega_{31}}+
{1\over \omega_{n'1}+\omega_{31}-\omega_{21}} \right)=1.018 
\end{eqnarray}
and 
\begin{eqnarray}
C_k &=&\sum_{n\ge 4}
{<p>_{n2}<p>_{n1}\over 3\omega_{31}}\Bigg[
\left( {\omega_{31}\over \omega_{n1}-\omega_{31}} \right)^k-
\left({-\omega_{31}\over \omega_{n1}+\omega_{31}-\omega_{21}} \right)^k \Bigg]
\nonumber\\
&+&\int_0^\infty dn' {<p>_{n'2}<p>_{n'1}\over 3\omega_{31}}\Bigg[
\left( {\omega_{31}\over \omega_{n'1}-\omega_{31}}\right)^k-
\left({-\omega_{31}\over \omega_{n'1}+\omega_{31}-\omega_{21}} \right)^k \Bigg],
\end{eqnarray}
for $k\ge 2$. Here, $f_{2s,3p}={2\over3}\omega_{32}^{-1}[<p>_{32}]^2=0.4349$ is the oscillator strength between the $2s$
and $3p$ states (e.g. Bethe \& Salpeter 1967).

Therefore, in frequency space the Raman scattering cross 
section is written as
\begin{eqnarray}
\sigma^{Ram}(\omega) &=& \sigma_T
\left({\omega_{31}\over\Delta\omega_2}
\right)^2
{\omega_{32}\over\omega_{31}} |C_0|^2
\left[1+\left({\omega_{21}\over\omega_{31}}+{2C_1\omega_{32}\over C_0\omega_{31}}\right)
\left({\Delta\omega_2\over\omega_{31}} \right)+\cdots\right]
\nonumber \\
&=&
\sigma_T
\left({\omega_{31}\over\Delta\omega_2}
\right)^2
{\omega_{32}\over\omega_{31}} |C_0|^2
\left[ 1-7.832\left({\Delta\omega_2\over\omega_{31}}\right) +\cdots\right].
\label{ram_betaf}
\end{eqnarray}
In wavelength space, we obtain
\begin{eqnarray}
\sigma^{Ram}(\lambda) &\simeq& \sigma_T
\left({\lambda_\beta\over\Delta\lambda_2}
\right)^2
{5\over 32} |C_0|^2
\Bigg[1+\left(-{17\over5}-2c_1\right) \left({\Delta\lambda_2\over\lambda_\beta}\right)
\nonumber \\
&+&\left( c_1^2+{44\over5}c_1 -{49\over5}+2c_2
\right)
\left( {\Delta\lambda_2\over\lambda_\beta } \right)^2 
\nonumber \\
&+&\left(
-2c_3-2c_1c_2-{27\over5}(c_1^2-2c_1+1+2c_2) \right)
\left({\Delta\lambda_2\over\lambda_\beta}\right)^3+\cdots \Bigg]
\nonumber \\
&=&
\sigma_T
\left({\lambda_\beta\over\Delta\lambda_2}
\right)^2
{5\over 32} |C_0|^2[1+5.223\times10^1(\Delta\lambda_2/\lambda_\beta)
\nonumber \\
&+&
9.103\times10^2(\Delta\lambda_2/\lambda_\beta)^2
-8.267\times10^3(\Delta\lambda_2/\lambda_\beta)^3+\cdots],
\label{ram_beta}
\end{eqnarray}
where the lower case coefficients $c_i$ are defined by
$ c_i=C_i/C_0$. The numerical values of these coefficients up to $c_5$ are
shown in Table~1.

Unlike the case for Rayleigh scattering near Ly$\beta$, the Raman scattering 
cross section shows redward asymmetry
with respect to the Ly$\beta$ center. This result can be traced 
to the fact that no contribution is made from
the $2p$ state, which played a dominant role 
in the case of Rayleigh scattering near Ly$\beta$. In the absence
of the $2p$ contribution, all the perturbing $p$
states are more energetic than Ly$\beta$.
This situation is exactly the same as the Rayleigh scattering around
Ly$\alpha$ resulting in redward asymmetry.

The ratio $r_b(\lambda)$ of the cross sections for Raman scattering
to Rayleigh scattering 
in the vicinity of Ly$\beta$ is given by
\begin{equation}
r_b(\lambda)={\sigma^{Ram}(\lambda)\over\sigma^{Ray}(\lambda)}=R_0+
R_1\left({\Delta\lambda_2\over\lambda_{\beta}}
\right)+R_2\left({\Delta\lambda_2\over\lambda_\beta}\right)^2+\cdots
\label{branching_eq}
\end{equation}
where the first three coefficients are explicitly 
$R_0=(5|C_0|^2/32)/(f_{13}/2)^2=2^{18}5^{-9}=0.1342$,  $R_1=11.26$ and $R_2=535.9$.
This result shows discrepancy with that provided by Yoo, Bak \& Lee (2002), 
in which there is an error in
their numerical calculation of the coefficients $A_3$ and $A_4$. 
The leading term can also be expressed as $R_0=(f_{2s,3p}/f_{13})
(\omega_{32}/\omega_{31})^2=0.1342$, which implies that the branching
ratio is determined by a combination of the oscillator
strength and the phase space volume factor represented by $\omega^2$. It is seen
that the dominant contribution is made by the phase
space volume available to scattered radiation.

In Fig.~1, we show the branching ratio $r_b(\lambda)$ in the 
neighborhood of Ly$\beta$. The solid line shows the result from
a direct numerical computation of the Kramers-Heisenberg formula.
The dotted line shows the linear fit and the dot-dashed line
shows the second order fit using Eq.~(\ref{branching_eq}). Because
the coefficient $R_2$ is large, the nonlinearity of $r_b(\lambda)$
is quite conspicuous in the figure. This behavior leads to a redward shift
in broad H$\alpha$ wings observed in young planetary nebulae 
and symbiotic stars which are also attributed to Raman scattering
of Ly$\beta$ (Jung \& Lee 2004).

\subsection{Total scattering cross section around Ly$\beta$}

In this subsection, we combine the results of previous subsections 
to provide the expansion of
the total scattering cross section around Ly$\beta$.
The total scattering cross section around Ly$\beta$ is the sum of
Eq.~(\ref{lybeq1}) and Eq.~(\ref{ram_betaf}), which is, 
to the first order of $\Delta\omega_2/\omega_{31}$, given by 
\begin{equation}
\sigma_{tot}(\omega) \simeq
\sigma_\beta \left(
{\omega_{31}\over\Delta\omega_2}\right)^2 
\left[
1+31.37\left({\Delta\omega_2\over\omega_{31}}\right) \right].
\end{equation}
Here, we introduce another parameter $\sigma_\beta$ defined by
\begin{equation}
\sigma_\beta=\sigma_T (f_{13}/2)^2[1+0.1342]=1.180\times10^{-27}{\rm\ cm^2}.
\end{equation}

In wavelength space,
we may combine Eq.~(\ref{lybeq2}) and Eq.~(\ref{ram_beta}) 
to express the total scattering cross section around Ly$\beta$ as
\begin{equation}
\sigma_{tot}(\lambda)\simeq \sigma_\beta \left(
{\lambda_{\beta}\over\Delta\lambda_2}\right)^2 \left[
1-24.63\left({\Delta\lambda_2\over\lambda_{\beta}}\right) \right].
\label{lybeq3}
\end{equation}

From this result, it is seen that 
 the Ly$\beta$ absorption profiles 
tend to shift blueward of the Ly$\beta$ line center. 
In Fig.~2, we show the total scattering cross section obtained from a numerical
evaluation of the Kramers-Heisenberg formula
around Ly$\alpha$ and Ly$\beta$ in wavelength space. 
The vertical axis shows the logarithm to the base 10
of $\sigma(\lambda)$ in units of cm$^2$. 

 In order to take a clear view of the asymmetry of the scattering cross section 
we plot the same quantities in Fig~3 as a function of the absolute value of the wavelength deviation. 
The cross sections redward of Ly$\alpha$ and Ly$\beta$ are shown with solid lines 
in the upper panel and lower panel, respectively.
The dotted lines show the cross sections blueward of Ly$\alpha$ and Ly$\beta$. 
The dotted lines are mirror images of the curves blueward of Ly$\alpha$ and Ly$\beta$ 
shown in Fig.~2. In Fig.~3, we see that red Ly$\alpha$ photons have larger scattering
cross section than blue counterparts 
and that the opposite is the case for Ly$\beta$.
 
In Fig.~4, we show the transmission coefficient $t(\lambda,N_{HI})$ defined by
\begin{equation}
t(\lambda, N_{HI}) \equiv 1-\exp[-\sigma(\lambda)N_{HI}]
\end{equation}
for various neutral hydrogen column densities. The upper panel is for Ly$\alpha$ 
and the lower panel is for Ly$\beta$. The solid line shows the result for  $N_{HI}=10^{20}{\rm\ cm^{-2}}$, 
the dotted line for $N_{HI}=10^{21}{\rm\ cm^{-2}}$,
and the dashed line for $N_{HI}=5\times 10^{21}{\rm\ cm^{-2}}$. 
In the case of $N_{HI}=5\times10^{21}{\rm\ cm^{-2}}$,
the asymmetry is quite noticeable in the scale shown in the figure.

\section{Asymmetry in the Absorption Profiles of Ly$\alpha$ and Ly$\beta$}

\subsection{Absorption center shift}

In this subsection, we investigate the shift of the absorption line center 
near Ly$\alpha$ and Ly$\beta$ as a function
of the neutral hydrogen column density. Denoting by $\lambda_c$ the line center wavelength of Ly$\alpha$ or Ly$\beta$, 
the scattering cross section is approximated to the first order of
dimensionless wavelength deviation from line center $x=\Delta\lambda/\lambda_c$ by a function 
\begin{equation}
f(x)={1\over x^2}+{a\over x}.
\end{equation}
The equation $f(x)=k>0$ has two solutions $x_1, x_2$, of which the mean is $x_m=a/( 2k)$. This implies that
the absorption center can be meaningfully defined when we fix the value of cross section. The sign of the
coefficient $a$ determines the direction of asymmetry, where a positive $a$ results in a red asymmetry.

Given a value of the H~I
column density $N_{HI}$ we define the
mean wavelength $\lambda_{m1}^\alpha$ of the two wavelengths $\lambda_{1,2}$ 
at which $\tau(\lambda_1)=\tau(\lambda_2)=\sigma(\lambda)N_{HI}=1$
around Ly$\alpha$ and in a  similar way we define $\lambda_{m1}^\beta$ for Ly$\beta$.
We also introduce $\lambda_{m2}^{\alpha}$ and $\lambda_{m2}^\beta$ 
as the mean value of the two wavelengths $\lambda_{1,2}'$ 
where we have $\tau(\lambda_{1}')=\tau(\lambda_2')=0.5$ around Ly$\alpha$
and Ly$\beta$, respectively. 
Corresponding to these wavelengths $\lambda_{m1}$, we define the velocity shift $\Delta V_1$ by the relation
\begin{equation}
\Delta V_1^{\alpha}\equiv  c(\lambda_{m1}^\alpha-\lambda_\alpha)/\lambda_\alpha
\end{equation}
for Ly$\alpha$ and in a similar way $\Delta V^\beta_1$ is defined  for Ly$\beta$. 
Here, $c$ is the speed of light.

In Table~2, we show the  values of $\lambda_{m1}^\alpha$ and $\lambda_{m2}^\alpha$ for various neutral 
hydrogen column densities.
Also in Table~3 we show the quantities corresponding to the Ly$\beta$ transitions. At $N_{HI}=10^{21}{\rm\ cm^{-2}}$
we obtain a redward center shift in the amount of $\Delta\lambda=+16{\rm\ km\ s^{-1}}$ for Ly$\alpha$ and 
and a blueward shift of $\Delta\lambda=-3.9{\rm\ km\ s^{-1}}$ for Ly$\beta$.

In Fig.~\ref{fig_cena},
we show $\Delta V_1^\alpha$ and $\Delta V_2^\alpha$ as a function of $N_{HI}$ 
in the cases of Ly$\alpha$ (upper panel) and $\Delta V^\beta_1$ and $\Delta V^\beta_2$ 
for Ly$\beta$ (lower panel). The dotted 
line shows a fit to the data, which implies that both $\Delta V_{1}^{\alpha,\beta}$ and 
$\Delta V_{2}^{\alpha,\beta}$ are proportional to $N_{HI}$.
The linear fit shown by the dotted line for Ly$\alpha$ in the figure is given by
\begin{equation}
\Delta V_1^\alpha\simeq 1.6 \left[{N_{HI}\over 10^{20} {\rm\ cm^{-2}}}\right] {\rm\ km\ s^{-1}},
\end{equation}
and similarly for Ly$\beta$ it is given by
\begin{equation}
\Delta V_1^\beta \simeq -0.39 \left[{N_{HI}\over 10^{20} {\rm\ cm^{-2}}}\right] {\rm\ km\ s^{-1}}.
\end{equation}

\subsection{Profile fitting by shifting the Lorentzian}

Another way of quantifying the asymmetry is provided by fitting 
the absorption profiles. In this subsection, 
we compare the transmission coefficient $t(\lambda,N_{HI})$ derived 
from the Kramers-Heisenberg formula with that obtained from the Lorentzian 
shifted by a finite amount. 
For simplicity, we fix the \ion{H}{1} column density 
$N_{HI}=5\times10^{21}{\rm\ cm^{-2}}$.
This procedure may illustrate an error estimate in determining the redshift 
of a DLA system with $N_{HI}=5\times10^{21}{\rm\ cm^{-2}}$.

In Fig.~6 we show the result for Ly$\alpha$. The solid line in each panel 
shows the transmission coefficient obtained 
from the Kramers-Heisenberg formula. The dotted line in the top panel shows 
the transmission coefficient 
from the Lorentzian function, which provides an excellent fit near the line 
center. However, a considerable deviation in the wing part is quite 
noticeable. 
With the dotted line in the bottom panel, we show the quantities obtained 
by shifting the Lorentzian redward by an amount of $+0.8{\rm\ \AA}$. 
Improvement of the fitting in wing parts is achieved 
only at the expense of a poor approximation near the line center. 
In the middle panel, we show the Lorentzian 
shifted by $+0.4{\rm\ \AA}$,
in which the quality of the fit is compromised between the previous two cases.

In the analysis by Lee (2003) the optimal wavelength shift was proposed by $+0.2{\rm\ \AA}$, 
which is 
smaller than $+0.4{\rm\ \AA}$ suggested in this work. This discrepancy is 
due to the fact that the fitting procedure 
in Lee (2003) was confined to a rather narrow interval 
of $|\Delta\lambda|<34{\rm\ \AA}$ excluding extreme wing parts.
The procedure of fitting a DLA profile using the shifted Lorentzian tends to overestimate the line center wavelength
of Ly$\alpha$ leading to corresponding overestimate of the redshift 
of the DLA. We note that Lee (2003) made 
a mistake in pointing out that the redshift would be {\it 'underestimated'}, 
which should be corrected to be {\it 'overestimated'}.

A similar analysis corresponding to Ly$\beta$ is shown in Fig.~7 
for the same neutral hydrogen column density $N_{HI}=5\times 10^{21}
{\rm\ cm^{-2}}$. In the figure, the dotted line in each panel 
shows the transmission coefficient from the Lorentzian function (top panel) 
and shifted Lorentzian functions (middle and bottom panels), 
whereas the solid line shows the exact transmission 
coefficient computed from the Kramers-Heisenberg formula. The amount 
of wavelength shift blueward of Ly$\beta$ is
$\Delta\lambda=-0.1{\rm\ \AA}$ and $-0.2{\rm\ \AA}$ for the middle and bottom panels, respectively. As in the case
of Ly$\alpha$ illustrated in Fig.~6, the unshifted Lorentzian gives an excellent fit to the core part of the absorption
profile whereas the bottom panel shows an improved fit to the wing parts 
with the loss of fitting quality at the core part.

In Fig.~8, we present the transmission coefficients using 
the Kramers-Heisenberg formula and the Lorentzian functions 
around Ly$\alpha$ and Ly$\beta$ in the wavelength interval 
between 980 \AA\ and 1400 \AA\ 
for a very thick \ion{H}{1} medium with $N_{HI}=5\times 10^{22}{\rm\
cm^{-2}}$. This kind of an extreme neutral hydrogen column density has been 
found toward the gamma ray burst 
GRB080607 (e.g. Prochaska et al. 2009). For comparison, we show the transmission coefficients obtained from the Lorentzian 
around Ly$\alpha$ and Ly$\beta$ 
by the dotted line and the dashed line, respectively.
In this highly thick medium, the deviation from the Lorentzian is quite severe due to the
contribution from higher order terms, which prevents one from obtaining satisfactory results by fitting the absorption
profiles by a Voigt  or equivalently a Lorentzian function. 

In particular,  in the wavelength range shown in Fig.~8, the local 
peak transmission is found at $\lambda_p=1062{\rm\ \AA}$, for which 
$t(\lambda, N_{HI}=5\times10^{22}
{\rm\ cm^{-2}})=0.0144$. 
However, the sum of two Lorentzian functions around Ly$\alpha$ and Ly$\beta$ admits a 
local maximum at $\lambda=1070{\rm \AA}$. 
This shows the inadequacy of using a Voigt function for fitting analyses 
in extended wing parts
in the case of very high column density systems. Furthermore, at this high $N_{HI}$, the blue wing region
of Ly$\alpha$ overlaps with that of the red Ly$\beta$ wing, for which 
full quantum mechanical formula
should be invoked for an accurate analysis.

\section{Summary and Discussion}

The Kramers-Heisenberg formula is expanded around Ly$\alpha$
and Ly$\beta$ in order to investigate the asymmetric deviation of the
scattering cross section. A redward asymmetry is seen around Ly$\alpha$ and
a blueward asymmetry is found around Ly$\beta$. For red Ly$\alpha$ photons
the perturbing transitions from $(n+n') p$ states $(n\neq2)$ 
provide a positive contribution to the scattering cross section because they are in the same
side as the $2p$ state in the energy space, resulting in red asymmetry.
In the case of Ly$\beta$, Rayleigh scattering contributes more than Raman
scattering by a factor 6.452. Raman scattering around Ly$\beta$ exhibits
a red asymmetry like Ly$\alpha$ because all the perturbing transitions lie
higher than the main transition. However, for Rayleigh scattering
around Ly$\beta$, the transition from $2p$ state is the dominant perturbing
transition which is less energetic than Ly$\beta$. This leads to a blue asymmetry
in $\sigma(\lambda)$ around Ly$\beta$. In an attempt to quantify these asymmetries
we compute the mean wavelengths for which the scattering optical depth becomes
a unity or one half for various values of \ion{H}{1} column density $N_{HI}$.
Also we fitted the transmission coefficients for given $N_{HI}$ by shifting
the Lorentzian function.

Peebles (1993) introduced the formula for resonance scattering cross section 
around Ly$\alpha$  
\begin{equation}
\sigma_P(\omega)={3\lambda_\alpha^2\Lambda^2 \over 8\pi}
{(\omega/\omega_\alpha)^4\over (\omega-\omega_\alpha)^2+(\Lambda^2/4)(\omega/\omega_\alpha)^6},
\label{peebles}
\end{equation}
which is often used in fitting wing profiles of Ly$\alpha$ 
(e.g. Miralda-Escude 1998).
In particular, the red damping wing of Ly$\alpha$ is essential to probe the partially 
neutral intergalactic 
medium expected around the end of cosmic reionization  
(Gunn \& Peterson 1965, Scheuer 1965, Mortlock et al. 2011).  
Neglecting the damping term in the denominator, this expression yields 
an expansion in frequency space
\begin{equation}
\sigma_P(\omega)\simeq \sigma_\alpha 
\left({\omega_{\alpha} \over  \Delta\omega}\right)^2
\left[1+4\left({\Delta\omega\over \omega_\alpha}\right) \right].
\end{equation}
In this expression, the coefficient of the first order term  is 4, which differs 
significantly from the correct value of $-1.792$. According to this
formula, the scattering cross section is larger in the blue part of Ly$\alpha$
than in the red part, which is incorrect. The discrepancy in the expansion 
may be traced to the approximation adopted in the derivation 
of Eq.~(\ref{peebles}), where the hydrogen atom is effectively treated 
as a two level system.

The Lorentzian or Voigt profile matches the Kramers-Heisenberg
profile excellently only in the core part. Therefore, the redshift will be 
measured reliably when the profile
fitting is more weighted toward deeply absorbed core part than far wing parts. 
With the accurate determination of the redshift and column density 
of the DLA, one may obtain reliable transmission coefficients using
the Kramers-Heisenberg formula or its first order approximation 
given in Eq.~(\ref{lya_wv}) and Eq.~(\ref{lybeq3}).

In an analysis of a quasar spectrum,  it is highly difficult 
to obtain the accurate continuum level due to intervening Ly$\alpha$ forest systems. 
Securing the quasar continuum level around the damped Ly$\alpha$ center 
with high precision is critical to verify the asymmetry presented in this work.
With the advent of extremely large telescopes in the near future equipped 
with a high resolution spectrometer
the accurate atomic physics will shed light on the physical conditions 
of neutral hydrogen reservoir
in the early universe.

\acknowledgments
The author is very grateful to the anonymous referee whose comments greatly 
improved the presentation of this paper.
This research was supported by the Basic Science Research Program 
through the National Research Foundation of Korea (NRF) funded 
by the Ministry of Education, Science and Technology (2011-0027069).

\appendix

\section{Angle Averaged Cross Section}

We show a detailed angular integration of the matrix element that constitute the
Kramers-Heisenberg formula. Because of the selection rule of the electric dipole 
interaction, the relevant states are $np$ and
$1s$ state in the case of Rayleigh scattering. In the case of Raman scattering 
relevant to  the interaction around Ly$\beta$, $2s$ state is also involved. 
However, as long as the anglular
and polarization average is concerned, the same calculation is performed.

A typical matrix element to be summed in the Kramers-Heisenberg formula is
\begin{equation}
M({\bf\hat e}^{(\alpha)},{\bf\hat e}^{(\alpha')},1s,I)=({\bf r}\cdot {\bf\hat e}^{(\alpha')})_{1s,I}
({\bf r}\cdot {\bf\hat e}^{(\alpha)})_{I,1s},
\end{equation}
where $I$ denotes an intermediate state. In particular, $I$ can be written as $I=|np,m>$, where
$m$ is the magnetic quantum number taking one of zero and $\pm 1$ in the case of a $p$-state.
The spherical harmonic functions with $l=1$ are explicitly defined by
\begin{equation}
Y_1^1(\theta,\phi)=-{1\over2}\sqrt{3\over 2\pi} \sin\theta e^{i\phi},
\quad
Y_1^{-1}(\theta,\phi)={1\over2}\sqrt{3\over 2\pi} \sin\theta e^{-i\phi},
\quad
Y_1^0(\theta,\phi)={1\over2}\sqrt{3\over \pi} \sin\theta e^{i\phi},
\end{equation}
from which we may set
\begin{equation}
{x\over r}=\sqrt{2\pi\over 3}(Y_1^{-1}-Y_1^1),\quad
{y\over r}=i\sqrt{2\pi\over 3}(Y_1^{-1}+Y_1^1),\quad
{z\over r}=2\sqrt{\pi\over 3}Y_1^0.
\end{equation}

Therefore given an intermediate state $I=|np,m>$, we have
\begin{equation}
M({\bf\hat e}^{(\alpha)},{\bf\hat e}^{(\alpha')},1s,I)
=<1s|(xe^\alpha_x+ye^\alpha_y+ze^\alpha_z)|np,m><np,m|
(xe^{\alpha'}_x+ye^{\alpha'}_y+ze^{\alpha'}_z)|1s>.
\end{equation}
The wavefunction $|np,m>$ is given by the product of the radial part $R_{n1}(r)$ and 
angular part $Y_1^m$, where as the $1s$ state is characterized by the radial part $R_{10}(r)$
multiplied by the trivial spherical harmonic $Y_0^0=1/\sqrt{4\pi}$. Therefore,
we have
\begin{eqnarray}
<1s|x|np,m>&=&<1s|r|np>\int d\Omega Y_0^0 \left[{x\over r}\right]Y_1^m
\nonumber \\
&=& <1s|r|np>(4\pi)^{-1/2}\sqrt{2\pi\over 3}[\delta_{m,-1}-\delta_{m,1}]
\nonumber \\
&=& <1s|r|np>{1\over\sqrt{6}}[\delta_{m,-1}-\delta_{m,1}].
\end{eqnarray}
Here, $\delta_{m,n}$ is the Kronecker delta and 
$<1s|r|np>=\int_0^\infty R_{10}(r) rR_{n1}(r) r^2 dr$ is the radial
expectation value between $1s$ and $np$ state.
In a similar way, for the operator $y$ and $z$ we have
\begin{eqnarray}
<1s|y|np,m> &=& <1s|r|np>{i\over\sqrt{6}}[\delta_{m,-1}+\delta_{m,1}]
\nonumber \\
<1s|z|np,m> &=& <1s|r|np>{i\over\sqrt{3}}\delta_{m,0}.
\end{eqnarray}
From this we note that
\begin{eqnarray}
M({\bf\hat e}^{(\alpha)},{\bf\hat e}^{(\alpha')},1s,I)
&=&|<1s|r|np>|^2\left[
{1\over\sqrt{6}}(\delta_{m,-1}-\delta_{m,1})e^{\alpha}_x
+{i\over\sqrt{6}}(\delta_{m,-1}+\delta_{m,1})e^{\alpha}_y \right.
\nonumber\\
&&\left.
+{1\over\sqrt{3}}\delta_{m,0}e^{\alpha}_z \right]
\times\left[
{1\over\sqrt{6}}(\delta_{m,-1}-\delta_{m,1})e^{\alpha'}_x \right.
\nonumber \\
&& \left.
+{-i\over\sqrt{6}}(\delta_{m,-1}+\delta_{m,1})e^{\alpha'}_y
+{1\over\sqrt{3}}\delta_{m,0}e^{\alpha'}_z \right]
\nonumber \\
&=&|<1s|r|np>|^2\left[{1\over6}\delta_{m,-1}(e^{\alpha}_x+ie^\alpha_y)
(e^{\alpha'}_x-ie^{\alpha'}_y) \right.
\nonumber \\
&& \left.
+{1\over6}\delta_{m,1}
(-e^\alpha_x+ie^\alpha_y)
(-e^{\alpha'}_x-ie^{\alpha'}_y)
+{1\over3}\delta_{m,0}e^\alpha_ze^{\alpha'}_z \right]
\end{eqnarray}

Given $np$ states, we sum over substates with $m=\pm1,0$ to obtain
\begin{equation}
\sum_{m=\pm 1,0}M({\bf\hat e}^{(\alpha)},{\bf\hat e}^{(\alpha')},1s,I)
={1\over3}|<1s|r|np>|^2 ({\bf e}^{\alpha}\cdot{\bf e}^{\alpha'}).
\end{equation}
As is well-known for Thomson scattering (e.g. pages 51 and 52 in Sakurai 1967), 
a numerical factor of $8\pi/3$ results from
averaging over polarization states for both incoming and outgoing radiation.

\begin{table}
\begin{tabular}{ccc}
\hline
Ly$\alpha$  &  Ly$\beta$  & Ly$\beta$ (Raman)  \cr
\hline
$a_1=-8.961\times 10^{-1}$ &  $b_1= 1.621 \times 10^1$    &  $c_1= -2.776\times10^1 $ \cr    
$a_2=-1.222\times 10^1$    &  $b_2= -4.299\times 10^2$    &  $c_2= -2.128\times10^2 $ \cr    
$a_3=-5.252\times 10^1$    &  $b_3= -2.176\times 10^3$    &  $c_3= -3.231\times10^3 $ \cr    
$a_4=-2.438\times 10^2$    &  $b_4= -6.005\times 10^4$    &  $c_4= -1.098\times10^5 $ \cr    
$a_5=-1.210\times 10^3$    &  $b_5= -8.414\times 10^5$    &  $c_5= -4.032\times10^6 $ \cr    
\hline
\end{tabular}
\caption{Expansion coefficients of Rayleigh scattering cross section around Ly$\alpha$
and Rayleigh and Raman scattering cross sections around Ly$\beta$. }
\end{table}

\begin{table}
\begin{tabular}{ccccc}
\hline
log $N_{HI}$  &  $\lambda_{m1}^\alpha$ (\AA)  &  $\Delta V_1^\alpha$  (km s$^{-1}$) & $\lambda_{m2}^\alpha$ (\AA) 
& $\Delta V_2^\alpha $ (km s$^{-1}$) \cr
\hline
   19.0 &      6.10E-04  &     0.151  &    1.34E-03  &    0.331 \cr    
   19.7 &      3.30E-03  &     0.813  &    6.47E-03  &    1.60 \cr    
   20.0 &      6.47E-03  &     1.60    &    1.32E-02  &    3.25 \cr    
   20.7 &      3.26E-02  &     8.04    &    6.51E-02  &   16.0 \cr    
   21.0 &      6.51E-02  &    16.0     &    1.30E-01  &   32.1 \cr   
   21.7 &      3.25E-01  &    80.2     &    6.51E-01  &  1.60E+02 \cr    
   22.0 &      6.51E-01  &1.60E+02 &    1.31          &  3.20E+02  \cr    
   22.7 &      3.25          &8.01E+02 &    6.48          & 1.60E+03 \cr 
\hline
\end{tabular}
\caption{Absorption center shifts around Ly$\alpha$ for various neutral hydrogen column densities. }
\end{table}

\begin{table}
\begin{tabular}{ccccc}
\hline
log $N_{HI}$  &  $\lambda_{m1}^\beta$ (\AA)  &  $\Delta V_1^\beta$  (km s$^{-1}$) & $\lambda_{m2}^\beta$ (\AA) 
&  $\Delta V_2^\beta $ (km s$^{-1}$) \cr
\hline
   19.0 &      -1.22E-04  &   -0.03568  &    -2.44E-04  &   -0.07136 \cr    
   19.7 &      -7.32E-03  &   -0.214  &    -1.34E-03  &    -0.392 \cr    
   20.0 &      -1.34E-03  &   -0.392  &    -2.69E-02  &   -0.785 \cr    
   20.7 &      -6.59E-03  &   -1.93   &     -1.34E-02  &   -3.92 \cr    
   21.0 &      -1.34E-02  &   -3.92   &     -2.67E-02  &   -7.81 \cr   
   21.7 &      -6.64E-02  &  -19.4    &     -1.32E-01  &  -38.5 \cr    
   22.0 &      -1.32E-01  &  -38.5    &     -2.60E-01   &  -75.9  \cr    
   22.7 &      -6.22E-01       &-1.82E+02&    -1.16      & -3.38E+02\cr 
\hline
\end{tabular}
\caption{Absorption center shifts around Ly$\beta$ for various neutral hydrogen column densities. }
\end{table}

\begin{figure}
\plotone{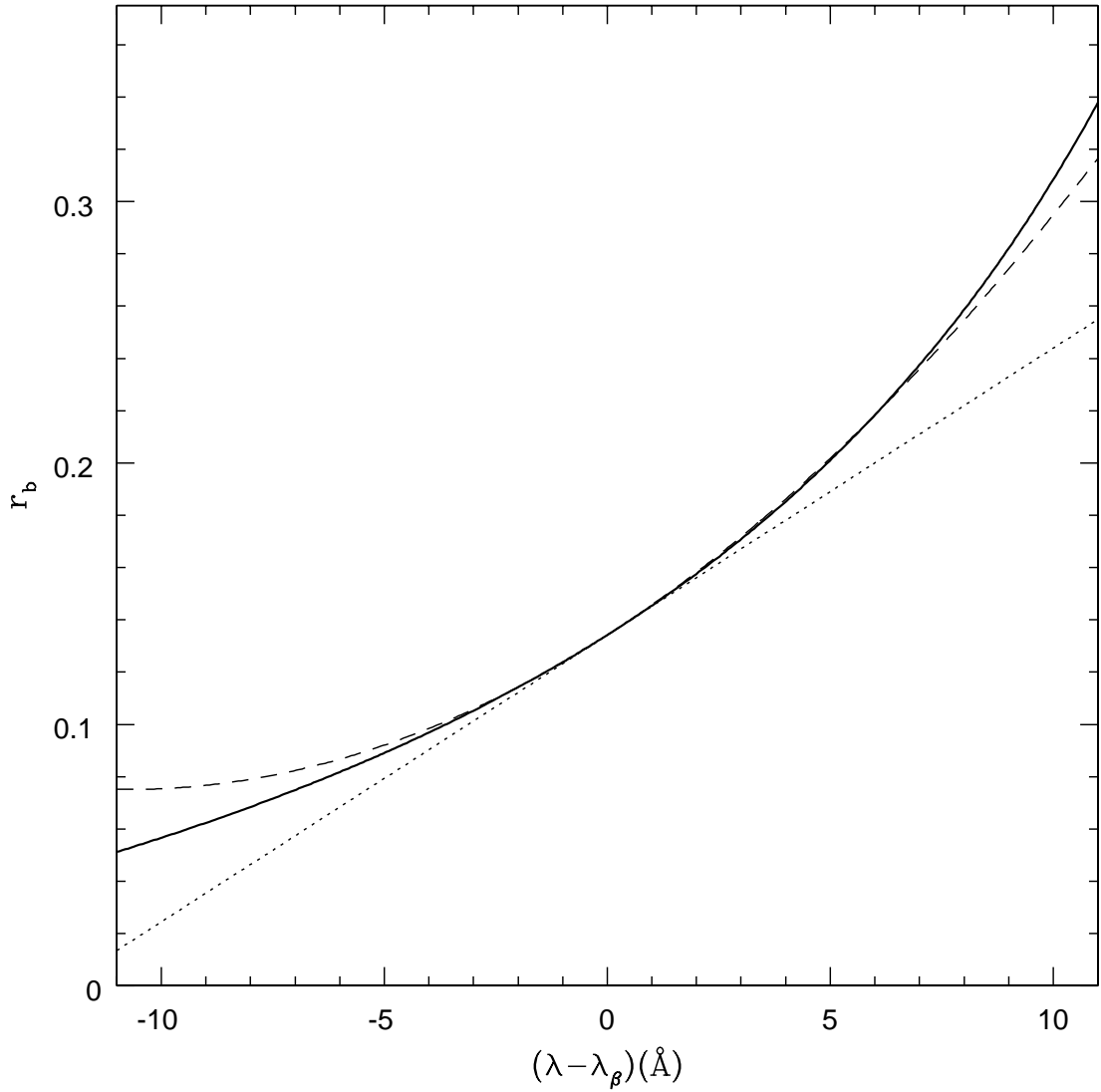}
\caption{Ratio of
cross sections of Raman scattering and Rayleigh scattering around Ly$\beta$.
The solid line shows the result from the full numerical calculation of the Kramers-Heisenberg formula. The dotted line shows the linear fit and
the dashed line shows the quadratic fit.
}
\label{branching_fig}
\end{figure}

\begin{figure}
\plotone{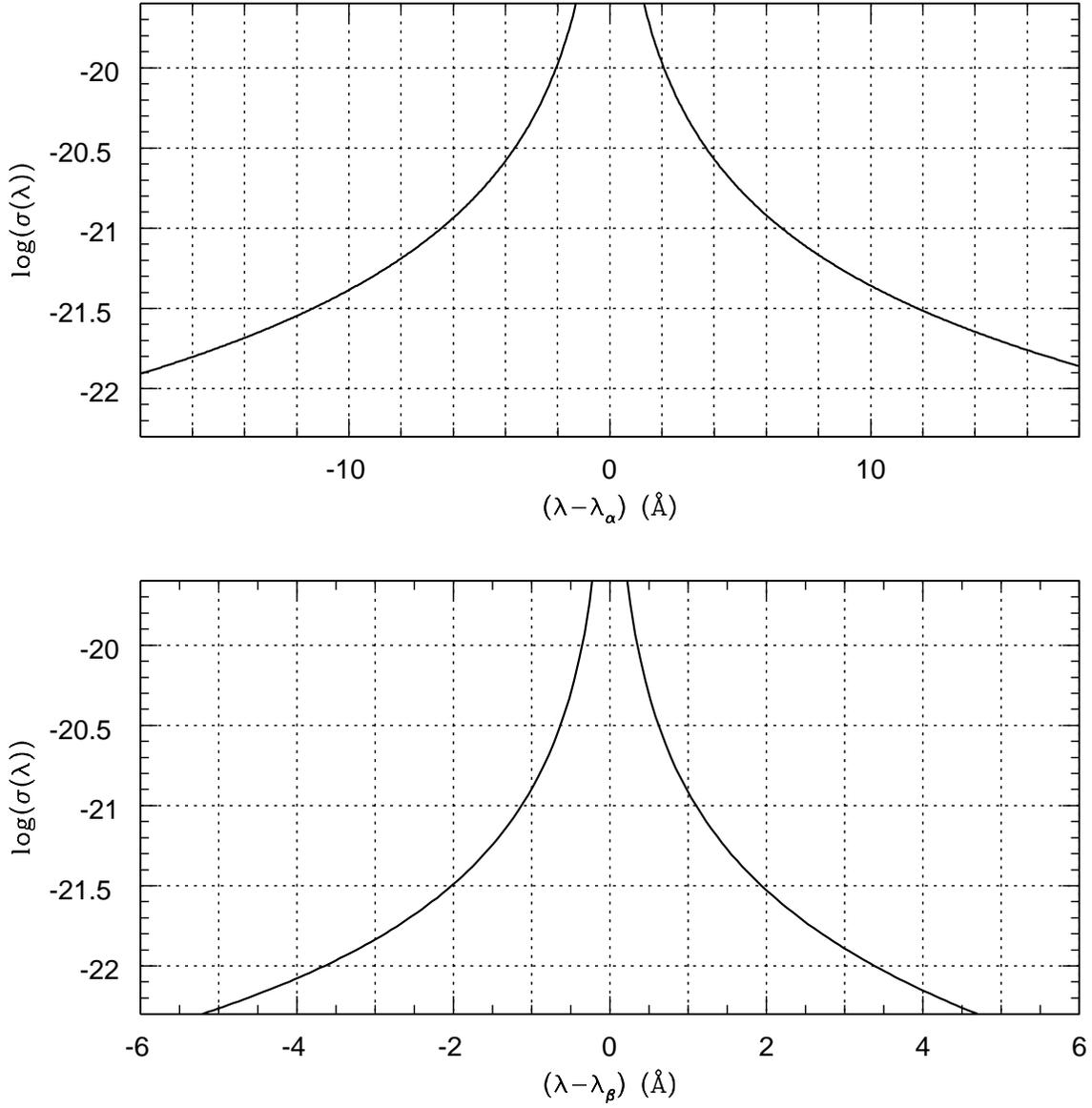}
\caption{Scattering cross section around Ly$\alpha$ (upper panel) and Ly$\beta$ (lower panel).
The horizontal axis shows the wavelength difference from the line center and the vertical axis represents
the logarithm of the cross section in units of cm$^2$.
}
\end{figure}

\begin{figure}
\plotone{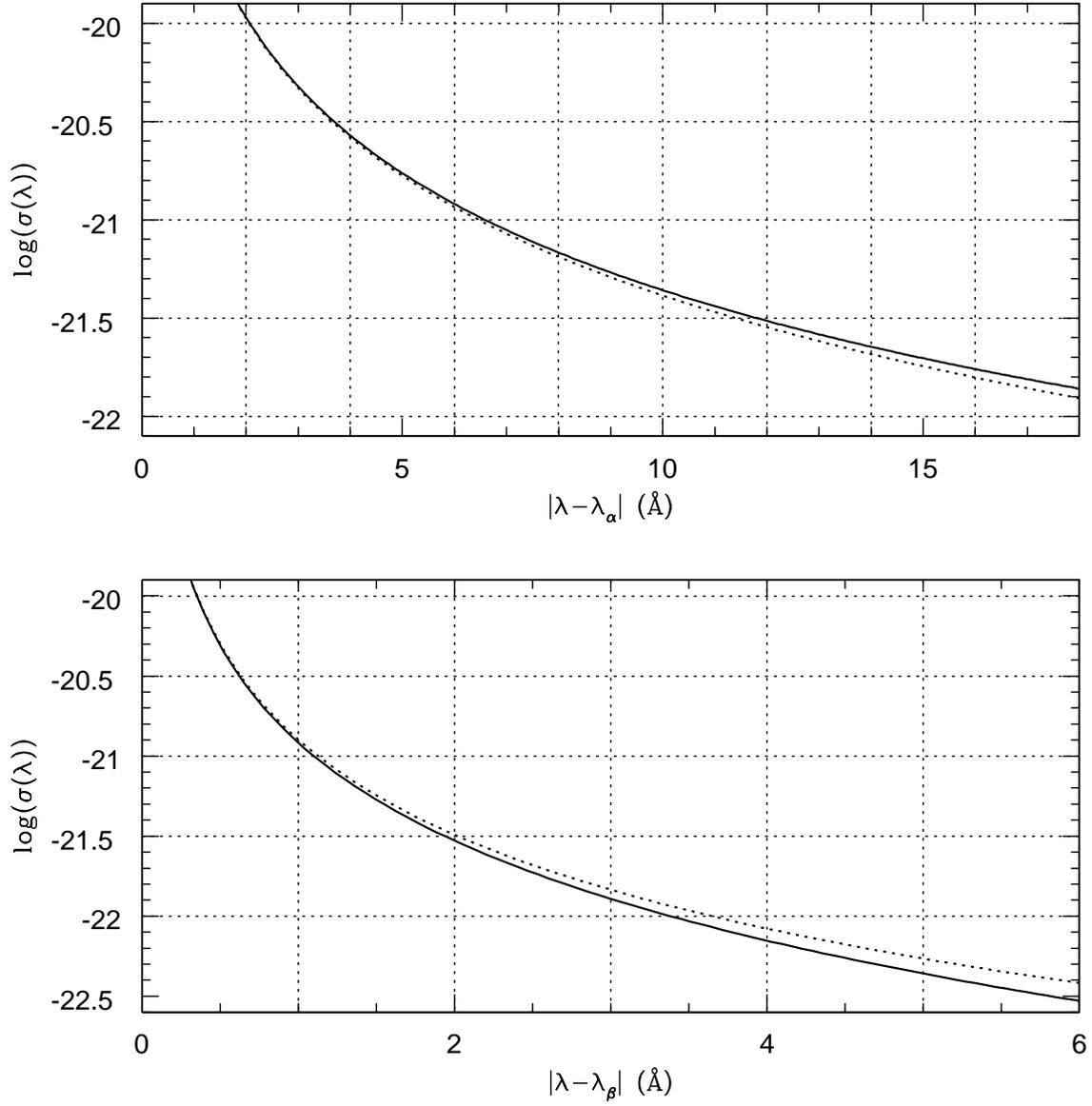}
\caption{Scattering cross section around Ly$\alpha$ (upper panel) and Ly$\beta$ (lower panel) as a function of
the absolute value of the wavelength difference.
The solid lines show the cross section redward of Ly$\alpha$ (upper panel) and Ly$\beta$ (lower panel). The
cross sections blueward of Ly$\alpha$ and Ly$\beta$ are shown in dotted lines in the upper panel and lower panel,
respectively.
}
\end{figure}

\begin{figure}
\plotone{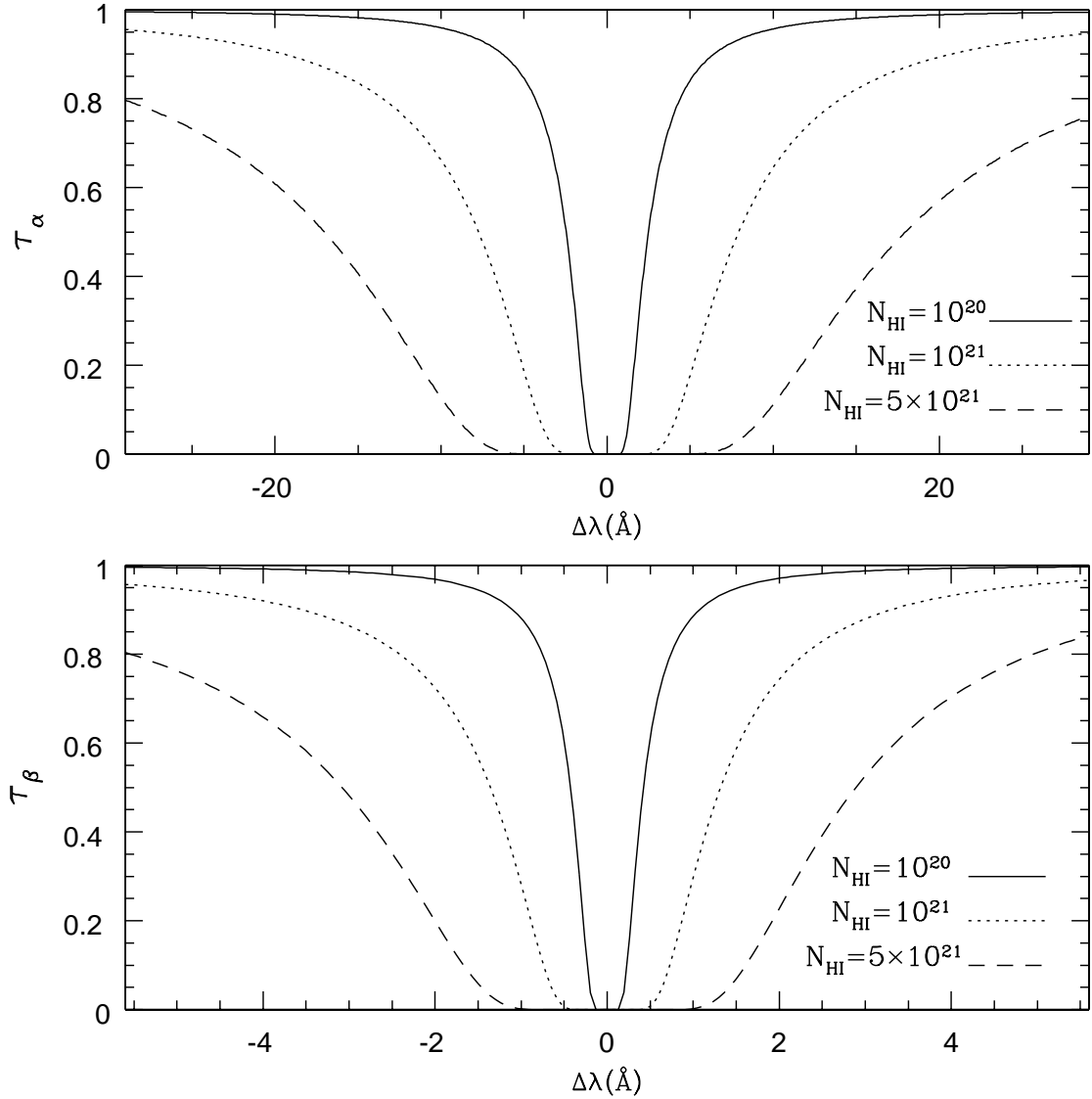}
\caption{Absorption profiles around Ly$\alpha$ (upper panel) and Ly$\beta$ (lower panel) for various neutral hydrogen column 
densities.
}
\label{fig_cena}
\end{figure}

\begin{figure}
\plotone{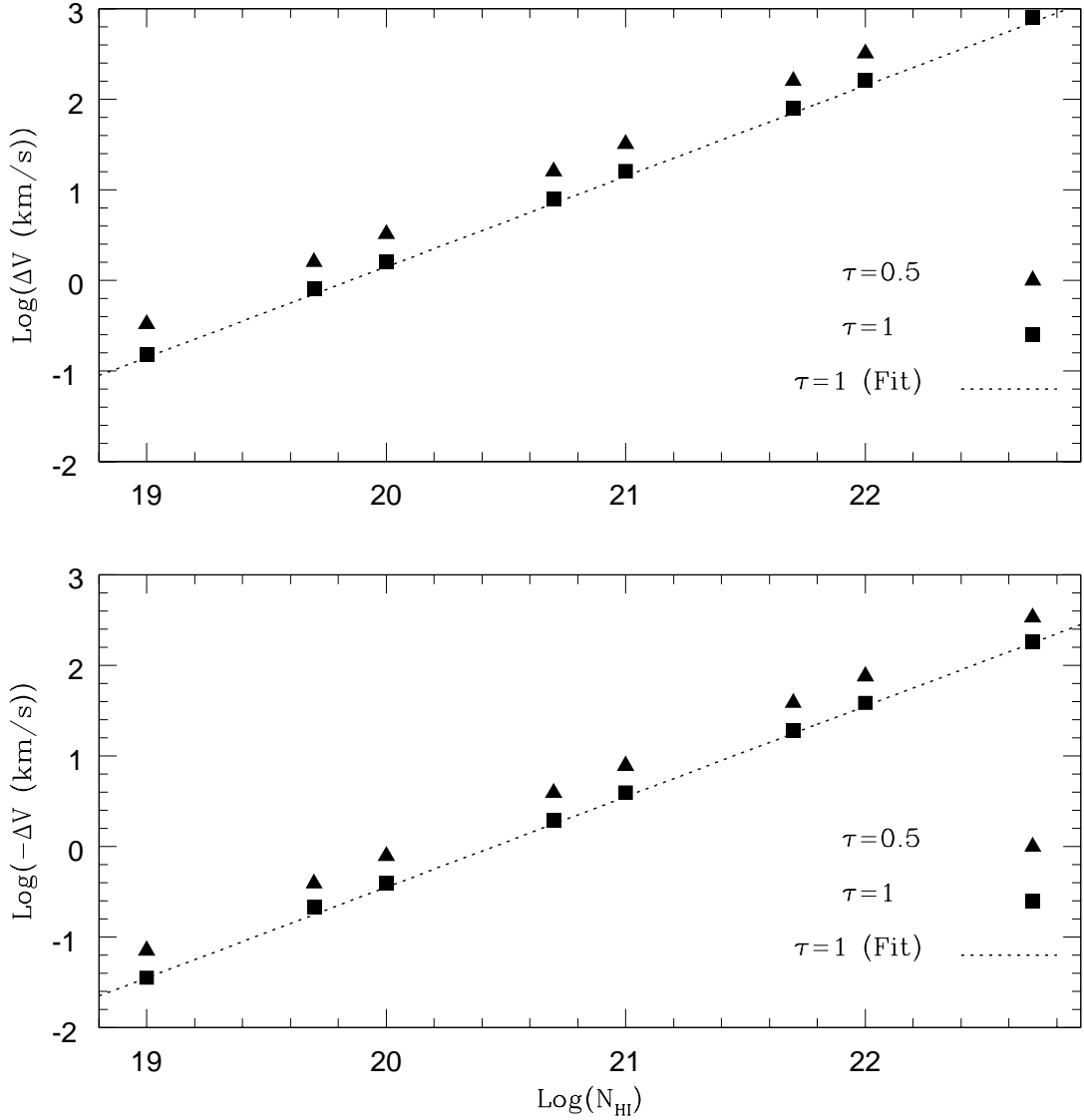}
\caption{Center shift of the absorption profile around Ly$\alpha$ (upper panel) and around Ly$\beta$ (lower panel). 
The horizontal axis shows the logarithm of the H~I
column density. The squares show the absorption center defined by the mean values of the wavelengths where $\tau(\lambda)=1$.
The triangles show the absorption center similarly defined by the condition $\tau(\lambda)=0.5$.
}
\label{fig_cena}
\end{figure}

\begin{figure}
\plotone{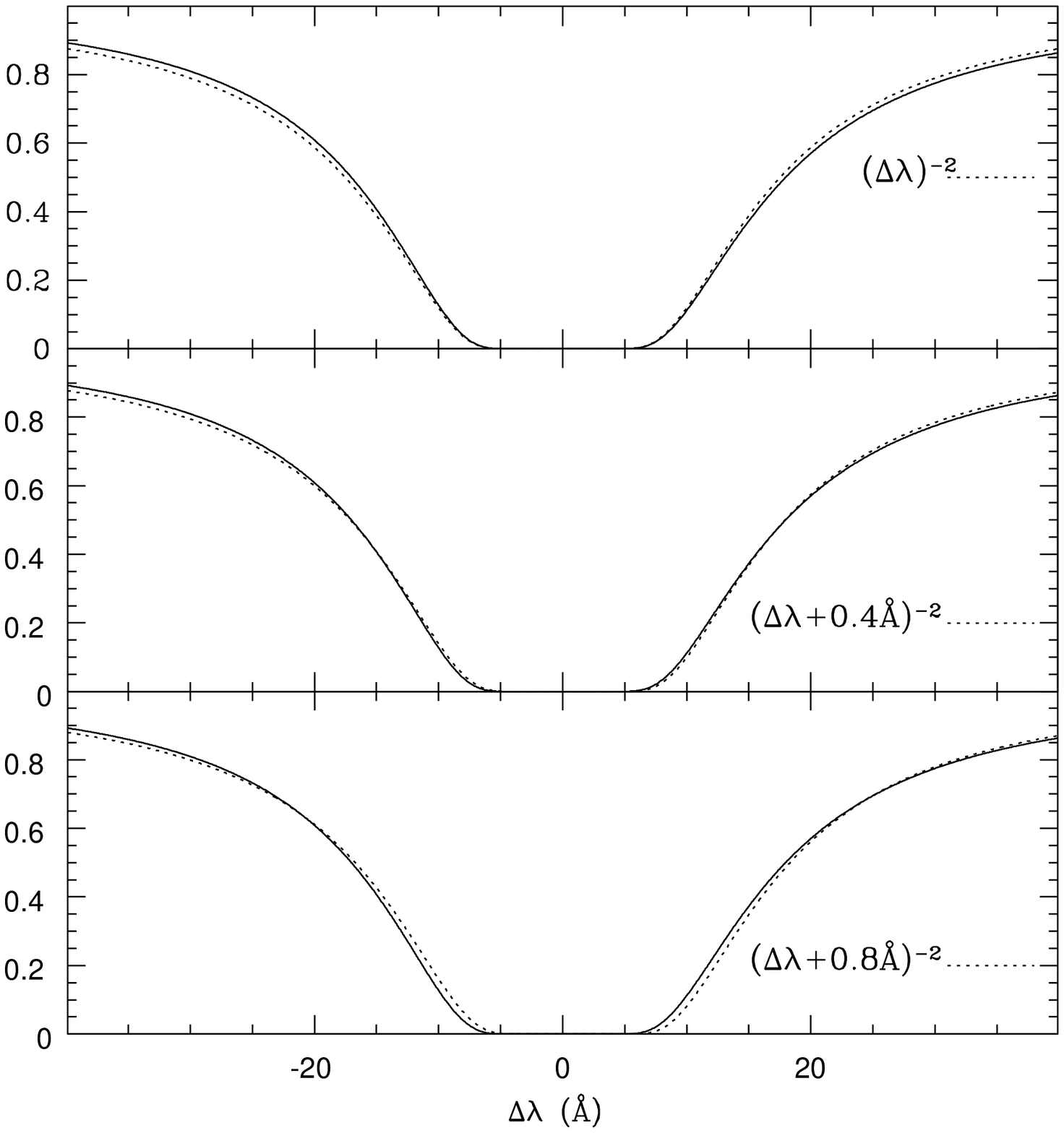}
\caption{Fit to the absorption profile around Ly$\alpha$ using shifted Lorentzian functions.
The solid line is the transmission probability for a neutral slab of hydrogen with $N_{HI}=5\times 10^{21}{\rm\ cm^{-2}}$.
}
\label{shifta}
\end{figure}

\begin{figure}
\plotone{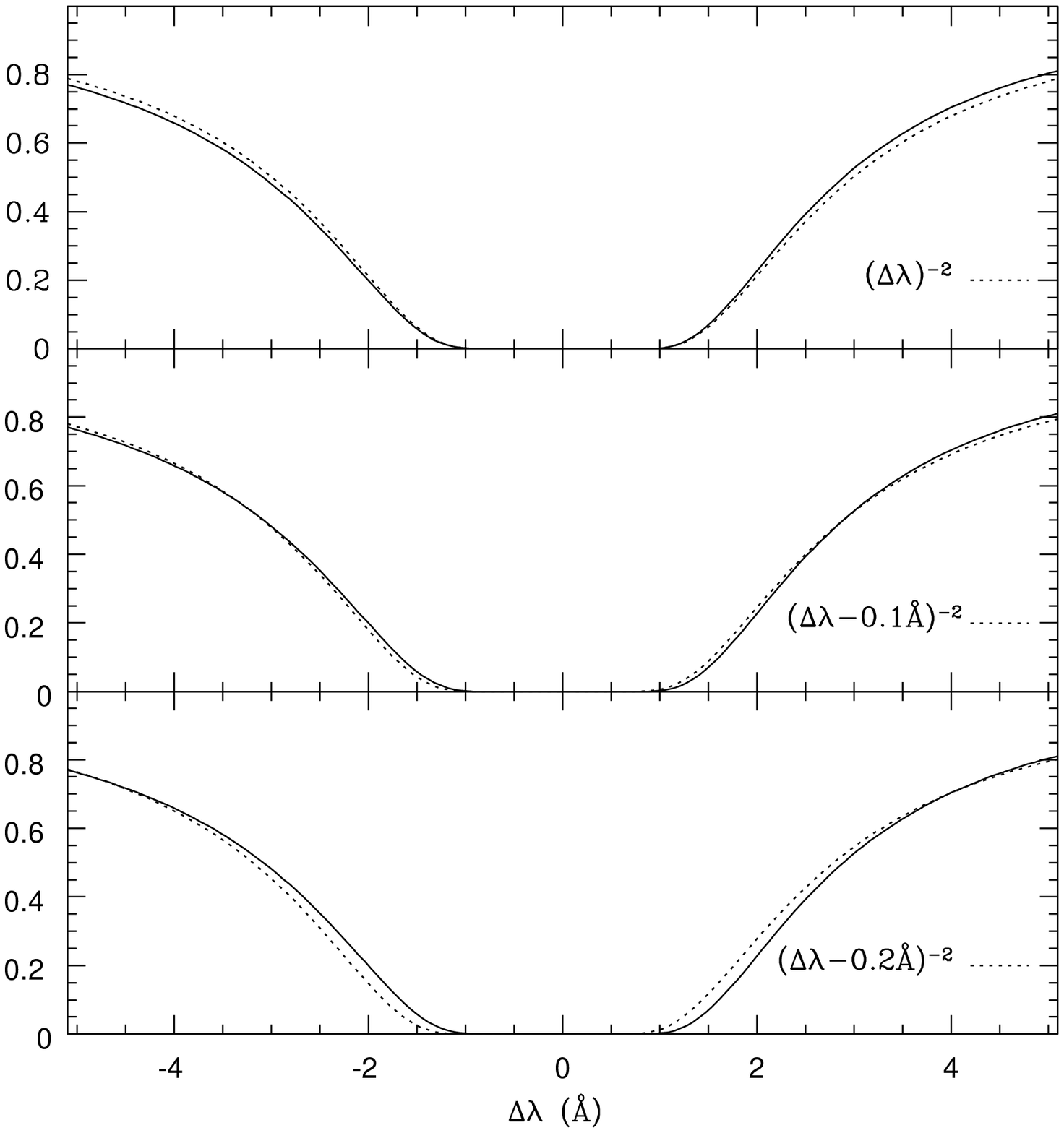}
\caption{Fit to the absorption profile around Ly$\beta$ using shifted Lorentzian functions.
The solid line is the transmission probability for a neutral slab of hydrogen with $N_{HI}=5\times 10^{21}{\rm\ cm^{-2}}$.
}
\label{shiftb}
\end{figure}

\begin{figure}
\plotone{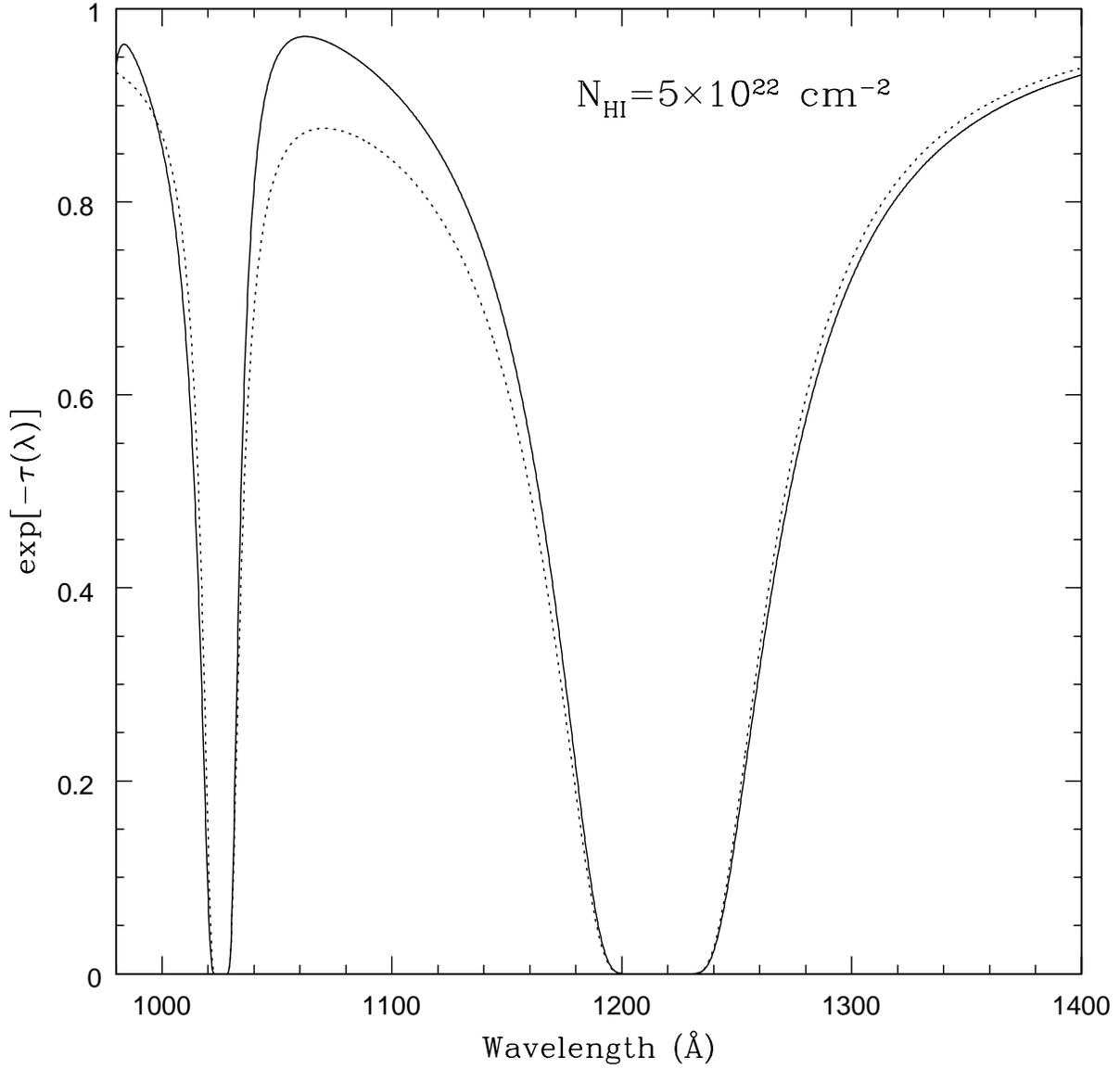}
\caption{Absorption profile in the wavelength region between 980 \AA\ and 1400 \AA\ for a slab of a neutral hydrogen column density
$N_{HI}=5\times 10^{22}{\rm\ cm^{-2}}$. 
The dotted line shows the transmission coefficient obtained from the sum of the two Lorentzian functions around Ly$\alpha$ and around Ly$\beta$.
}
\label{shiftb}
\end{figure}


\begin{thebibliography}{}
\bibitem[Arrieta \& Torres-Peimbert 2003]{arr03} Arrieta, A., Torres-Peimbert, S., 2003, ApJS, 147, 97
\bibitem[Berestetskii, Lifshitz, \& Pitaevskii, 1971]{ber71}
Berestetskii, V.B., Lifshitz, E.M., \& Pitaevskii, L.P., 1971,
Relativistic Quantum Mechanics, Pergamon Press
\bibitem[Bethe \& Salpeter 1967]{bet67} Bethe, H. A. \& Salpeter, E. E.  1967,  
Quantum Mechanics of One and Two Electron Atoms, Academic Press Inc., New York
\bibitem[Birriel 2004]{bir04} Birriel, J., 2004, ApJ, 612, 1136
\bibitem[Calura, Matteucci \& Vladilo 2003]{cal03} Calura, F., Matteucci, F., Vladilo, G., 2003,  MNRAS, 340, 59
\bibitem[Curran et al. 2002]{cur02} Curran, S. J., Webb, J. K., Murphy, M. T., Bandiera, R., Corbelli, E., Flambaum, V. V.,
2002, PASA, 19, 455
\bibitem[Gunn \& Peterson 1965]{gun65} Gunn, J. E., Peterson, B. A., 1965,
ApJ, 142, 1633
\bibitem[Isliker, Nussbaumer \& Vogel 1989]{isl89}Isliker, H., Nussbaumer, H., \& Vogel, M., 1989, A\& A,  219, 271 
\bibitem[Jung \& Lee 2004]{jun04} Jung, Y.-C., Lee, H.-W., 2004, MNRAS, 350, 580
\bibitem[Khare et al. 2012]{kha12} Khare, P., vanden Berk, D., York, D. G., 
Lundgren, B., Kulkarni, V. P., 2012, MNRAS, 419, 1028
\bibitem[Kim et al. 2011]{kim11} Kim, J., Park, C., Rossi, G., Lee, S. M., Gott III, R., 2011, Journal of the Korean
Astronomical Society, 44, 217
\bibitem[Kulkarni et al. 2012]{kul12} Kulkarni, V. P., Meiring, J., Som, D., P\'eroux, C., York, D. G., Khare, P., Lauroesch, J. T.,
2012, ApJ, 749, 176
\bibitem[Lee 2000]{lee00} Lee, H. -W., 2000, ApJ, 541, L25
\bibitem[Lee 2003]{lee03} Lee, H. -W., 2003, ApJ, 594, 637
\bibitem[Lee 2012]{lee12} Lee, H. -W., 2012, ApJ, 750, 127
\bibitem[Lee \& Lee 1997]{lee97} Lee, H. -W., Lee, K. W., 1997, MNRAS, 287, 211
\bibitem[Lee et al. 2006]{lee06} Lee, H. -W., Jung, Y. -C., Song, I. -O.,
Ahn, S. -H., 2006, ApJ, 636, 1045
\bibitem[Meiksin 2009]{mei09} Meiksin, A. A., 2009, Reviews of Modern Physics, 81, 2405 
\bibitem[Merzbacher 1961]{mer61} Merzbacher, E. 1970, Quantum Mechanics,
Wiley, New York
\bibitem[Miralda-Escud\'e 1998]{mir98} Miralda-Escud\'e, J., 1998, \apj,
501, 15
\bibitem[Mortlock et al. 2011]{mor11} Mortlock, D. J. et al. 2011, Nature,
 474, 616
\bibitem[Noterdaeme et al. 2012]{not12} Noterdaeme, P., Lauresen, P., Petitjean, P., Vergani,  S. D., Maureira, M. J.,
Ledoux, C., Fynbo, J. P. U., L\' opez, S., Srianand, R., 2012, A\& A, 540, A63
\bibitem[Nussbaumer, Schmid, \& Vogel 1989]{nus89}  Nussbaumer, H., Schmid, H. M.\& Vogel, 
M., 1989, A\&A, 221, L27 
\bibitem[Peebles 1993]{peb93} Peebles, P. J. E., 1993,
Principles of Physical Cosmology, Princeton University Press, Princeton
\bibitem[Prochaska et al. 2005]{pro05} Prochaska, J. X., Herbert-Fort, S.,  Wolfe, A. M., 2005, ApJ, 635, 123
\bibitem[Prochaska et al. 2009]{pro09} Prochaska, J. X. et al., 2009, ApJ, 691,
L127
\bibitem[Rafelski et al. 2012]{raf12} Rafelski, M., Wolfe, A. M., Prochaska, J. X., Neeleman, M., Medez, A. J., 2012, ApJ,
755, 89
\bibitem[Rauch 1998]{rau98} Rauch, M., 1998, ARA\&A, 36, 267
\bibitem[Rybicki \& Lightman, 1979]{ryb79} Rybicki, G. B., Lightman, A. P., 1979,
Radiative Processes in Astrophysics,  Wiley-Interscience, New York
\bibitem[Sakurai 1967]{sak67} Sakurai, J. J., 1967, Advanced Quantum Mechanics,
Addison-Wesley Publishing Company, Reading, Massachusetts
\bibitem[Saslow \& Mills 1969]{sas69} Saslow, W. M., Mills, D. L. 1969, Physical Review, 187, 1025 
\bibitem[Scheuer 1965]{sch65} Scheuer, P. A. G. 1965, Nature, 207, 963
\bibitem[Schmid 1989]{sch89} Schmid, H. M. 1989,A\& A,  211, L31 
\bibitem[Wolfe et al. 1986]{wol86} Wolfe, A., Turnshek, D. A., Smith, H. E., Cohen, R. D., 1986, 
ApJS, 61, 249
\bibitem[Wolfe et al. 2005]{wol05} Wolfe, A., Gawiser, E., Prochaska, J. X., 2005, 
ARA\&A, 43, 861
\bibitem[Yoo et al. 2002]{yoo02} Yoo, J. J., Bak, J.-Y., Lee, H.-W., 2002, MNRAS, 336, 467
\end{thebibliography}
\end{document}